\documentstyle[preprint,aps]{revtex}
\begin{document}
\newcommand{\ch}{\mbox{$\chi^{(2)\;}$}}
\newcommand{\cw}{\mbox{$\chi^{(2)}(\omega)\;$}}
\newcommand{\cwev}{\mbox{$\chi^{(2)}_{even}(\omega)\;$}}
\newcommand{\cwodd}{\mbox{$\chi^{(2)}_{odd}(\omega)\;$}}
\newcommand{\cwt}{\mbox{$\chi^{(2)}(\omega ,T)\;$}}
\newcommand{\cz}{\mbox{$\chi^{(2)}_{zzz}(\omega ,T)\;$}}
\newcommand{\Iw}{\mbox{$I_{2\omega}(\omega)\;$}}
\newcommand{\iwt}{\mbox{$I_{2\omega}(\omega ,T)\;$}}
\newcommand{\epsw}{\mbox{$\epsilon(\omega ) \;$}}
\newcommand{\ewt}{\mbox{$\epsilon(\omega ,T) \;$}}
\newcommand{\Lam}{\mbox{$\Lambda $} }
\newcommand{\hw}{\mbox{$\hbar\omega $} }
\newcommand{\hhw}{\mbox{$2 \hbar\omega $} }

\title{Theory for Spin-Polarized Oscillations in Nonlinear
Magneto-Optics due to Quantum Well States}
\author{ T. A. Luce, W. H\"ubner, and K. H. Bennemann}
\address{ Institute for Theoretical Physics,
 Freie Universit\"at Berlin, Arnimallee 14, D-14195 Berlin , Germany}

\date{\today}
\maketitle

\begin{abstract}
\leftskip 54.8pt
\rightskip 54.8pt
Using an electronic tight-binding theory we calculate the nonlinear
magneto-optical response from an x-Cu/1Fe/Cu(001) film as a function
of frequency and Cu overlayer thickness (x=3$\ldots$25). We find very
strong spin-polarized quantum well oscillations in the {\em nonlinear}
magneto-optical Kerr effect (NOLIMOKE). These are enhanced by the
large density of Fe $d$ states close to the Fermi level
acting as intermediate states for frequency doubling. In good
agreement with experiment we find two oscillation periods of 6-7 and 11
monolayers the latter being more pronounced.

\end{abstract}


\narrowtext


%
The magnetism of low-dimensional metallic structures such as surfaces,
thin films, and multilayer sandwiches has recently become an exciting new
field of research and applications~\cite{leo}. In particular, thin magnetic
films and multilayers exhibit a rich variety of properties not previously
found in bulk magnetism such as enhanced or reduced moments~\cite{weller},
oscillatory exchange coupling through nonmagnetic
spacers~\cite{grunberg,pierce,parkin},
giant magnetoresistance~\cite{fert,mertig}, and the
reorientation of the magnetic easy axis upon thickness and
temperature variation~\cite{bader,weller1,allenspach,schulz}.
Especially the observation
of spin-polarized quantum well states (QWS)~\cite{ortega,himpsel91,rader,garrison} in Cu/Co(001) has
attracted a great deal of attention. It has become clear that quantum
well states are indeed responsible for the important
oscillatory behavior of the exchange coupling of ferromagnetic thin films via
nonmagnetic spacers~\cite{edw_mathon91,schilfgaarde93}. Presently mainly
photoemission (PE)
and inverse photoemission (IPE)~\cite{ortega,himpsel91,rader,garrison}
have been used to identify QWS effects. Very recently a possible
connection between thickness dependent changes in NOLIMOKE and
QWS~\cite{wierengaprl95} has been proposed.  

It is the goal of this Letter to show that also {\em nonlinear}
optics, in particular NOLIMOKE, is a new sensitive tool for studying QWS. We 
find very interesting structure in the NOLIMOKE signal due to
particular transitions in {\bf k}-space. This is very remarkable since
it indicates that NOLIMOKE is able to detect very sensitively
{\bf k}-dependent structures. This new effect seems to be of general
interest for the 
physics of nonlinear optics and its relationship to the underlying
electronic structure. Note, this is not the case for linear optics,
since there the contribution of the Drude term of the dielectric
function creates a strong background of transitions from all 
{\bf k}-directions. Nonlinear optics, in contrast to linear optics, is
able to give angle-resolved
information about the underlying electronic structure. 
We demonstrate this by extending previous work on the Fe/Cu(001)
bilayer system \cite{puluhue96}
to the sandwich system x-Cu/Fe/Cu(001) where the layer
number x is varied between 3 and 25.
Thus we calculate the magnetic intensity contrast 
$\Delta I_{2\omega } = \frac{I_{2\omega}({\bf M})-I_{2\omega}({\bf -
    M})}{I_{2\omega}({\bf M})+I_{2\omega}({\bf - M})}$ 
of NOLIMOKE for these 
systems and find very large quantum well oscillations, originating
from particular transitions in {\bf k}-space.

In view of the electronic structure presented in Fig.~\ref{band}, a simple
physical picture already explains the occurrence of quantum well
oscillations in NOLIMOKE. One gets the main peak of the multilayer
system, since for 11 layer and multiples of this the marked
transitions (a) between Cu $d$-bands and the quantum well states as final
states become resonant at $2 \hbar\omega$. Obviously this causes an
oscillation with a period of 11 monolayers (ML). Correspondingly the period
of 6-7 ML results from the marked transitions (b) in Fig.~\ref{band}. Also it
becomes clear that the spin polarization of the intermediate Fe states will
cause a magnetization dependence and in particular a shifting of the
peak for the magnetization direction {\bf M} to lower periods. For the situation sketched in
Fig.~\ref{band} the {\bf k}-selectivity becomes immediately obvious since
unoccupied final states are necessary for a contribution to the SHG
yield. 

To verify these physical expectations we performed calculations using
our previous theory \cite{hub89,hub_bohm94,luhue96} to evaluate the SHG
intensity \Iw for opposite magnetization directions. 
Employing an electronic theory for
both the nonlinear susceptibility and the dielectric function and
separating \cw into even and odd parts under magnetization reversal
\cwev and \cwodd, we get for the SHG 
yield within the electric dipole approximation 
for the polar geometry (i.e. {\bf M} normal to the surface) \cite{hue95}
\begin{eqnarray}
\lefteqn{I_{2\omega}({\bf \pm M})\;=\;\mid 2i \mid
  E_{0}^{(\omega)}\mid^{2}\{ \cwev A_p [
2 F_c f_c f_s}   \nonumber \hspace{10mm}\\
&& + N^2 F_s (f_c^2 + f_s^2)]t_p^2 \cos^2 \varphi \cos \Phi \}   \nonumber \\
&&+ \{A_s [\cwev 2 f_s t_p t_s \cos \varphi \sin \varphi \nonumber \\
&&\pm \cwodd 2 f_c f_s t_p^2 \cos^2 \varphi ] \}sin \Phi \mid^2 
\end{eqnarray}
Here $\varphi $ and $\Phi$ denote the angle of polarization
of the incident light and the outgoing second harmonic light and were chosen
as  $\varphi = 
0^{\circ}$ ($p$-polarized) and  $\Phi = 75^{\circ}$. The linear
amplitudes $A_p$ and the transmission and Fresnel factors $t_{p,s}$, 
$f_{c,s}$ and $F_{c,s}$ are derived from the dielectric function
$\epsilon(\omega)$.
Note, the nonlinear susceptibility tensor
\cw is material specific via the electronic 
bandstructure, and so are the linear dielectric 
function $\epsilon (\omega)$
and the indices of refraction $n$ and $N$. To simplify our
calculation, we assume constant matrix elements,  
which are fitted to the linear dielectric function
$\epsilon(\omega)$ \cite{luhue96,hub}. This approximation 
is reasonable because 
the {\bf k}-dependence of the matrix elements is expected to
become less important in two dimensions due to the shrinking of the
$d$-band width for the reduced coordination number and also due to the
occurrence of additional allowed optical transitions \cite{luhue96}.
 Selection rules excluding dipole transitions of the type
 $\langle \Psi_{m=\pm 2} | r | \Psi_{m=\pm 0}\rangle$
 were taken into account~\cite{hub_bohm94}. To compare with
 experiment~\cite{straub95,kirilyuk96}, 
 we choose 1.61 eV as incident photon energy. 
A normalization with respect to the Cu layer number has to be
performed to take the interface sensitivity of SHG into account in
order to make the nonlinear response comparable for various
film thicknesses. Therefore, we
divided \Iw by the layer number, ensuring that for a band structure
without dispersion the response is identical for all layer numbers.
To calculate $\epsilon(\omega)$ and \cw from the electronic band
structure of the x-Cu/1Fe/(001)Cu system we use a Cu bulk
Hamiltonian (thus depending on $k_x$, $k_y$, $k_z$; $k_z = {\bf
  k}^{\perp}$ is 
perpendicular to the layers ) combined with a Fe
monolayer. The Hamiltonian is calculated within the Combined
Interpolation Scheme \cite{ehren-hodges}, the parametrisation is
according to Fletcher and Wohlfahrt
\cite{fletcher-wohl}. The parameters for the Cu bulk
bandstructure are taken from \cite{smith78}, for the Fe monolayer they
have been achieved from a fit to an {\em ab initio} calculation
\cite{pusto}. Of course, in $\Gamma$ - $X$ direction there is no
dispersion of 
the Fe monolayer band structure. We are evaluating the SHG
response at ($k_x$, $k_y$) = (0,0)), since for the (001) direction
the high density of states due to the extremal Fermi surface diameter
(caliper) at ${\bf k}_{\|} = (0,0)$, which give the QW period from
Ruderman-Kittel-Kasuya-Yoshida (RKKY) calculations~\cite{brunoPRL}
dominates the output\cite{footnote,ortega_himpsel93}. The 
{\bf k}-summation is performed over {\bf k}-points along the ${\bf
  k}^{\perp}$ direction. 

Note, due to the two resonance denominators  of \cw \cite{hub} it is not
necessary for 
the intermediate state to be unoccupied to give a contribution to
\cw. It is sufficient if the final state (usually a Cu {\em s} state) is 
unoccupied. As a consequence, a high density of intermediate states
leads to a large number of contributing terms to \cw, thus enhancing the SHG response. 
In our electronic structure, this amplification is caused by the spin
polarized Fe {\em d} states. Since at least
one of the 
three states involved in a nonlinear transition must be unoccupied to
give a contribution to the SHG yield, the QWS above $E_F$ are of great
importance as final states for the NOLIMOKE signal.
The QWS result from the confinement of the electrons in thin films,
causing an equally spaced discretization in ${\bf k}^{\perp}$-direction,
whereby the number of {\bf k}-points equals the number of layers.
Clearly this discretization of the {\bf k}-values affects the SHG 
intensity since photon transitions are limited to these 
distinct ${\bf k}^{\perp}$ points.

For the fundamental period \Lam  of intensity oscillations as
observed in photoemission, transitions at ${\bf k}^{\perp}$-vectors are
decisive, for which (for
Cu bulk) the $s$-band crosses the Fermi surface. If the layer 
number increases, such unoccupied ${\bf k}^{\perp}$-states at the Fermi
surface occur if the layer number equals 
$m \cdot{\bf k}^{\perp}_{BZ}/({\bf
k}^{\perp}_{BZ}-{\bf k}^{\perp}_{F})$, with 
$m=1,2,3 ...$. Then 
these new unoccupied $s$-states permit additional transitions, and the
optical response increases. The ratio
${\bf k}^{\perp}_{BZ}/({\bf k}^{\perp}_{BZ}-{\bf k}^{\perp}_{F})$ 
gives the fundamental period \Lam. Obviously, since optical transitions
may occur to all states above
$E_F$, this period only marks a lower limit of possible oscillation
periods and is not as strict as for (I)PE experiments, and depends on
the photon energy and the position of 
the initial bands. In particular for the nonlinear response, due to
the additional degree of freedom and due to \hhw resonances the SHG
intensity increases and new (larger) oscillation periods occur. Although
every period longer than the fundamental one may occur in the 
SHG spectrum if $d$ states allow for resonances with unoccupied QWS at
a ${\bf k}^{\perp}$ vector 
between ${\bf k}^{\perp}_{F}$ and ${\bf k}^{\perp}_{BZ}$, the period \Lam obtained from photoemission
experiments and the doubled period $2 \Lam$ have an outstanding
importance. If a QWS allows for a SHG signal with period $2 \Lam$,
there is a QWS at ${\bf k}_F$ too, both resonant transitions (a) and
(b) indicated in
Fig.~\ref{band} contribute to the SHG signal at layer thickness
$n\cdot 2\Lam$, thus enlarging the SHG amplitude at $m\cdot
2\Lam$. Due to interferences of   
the various transitions, this enhancement is not compensated by the
performed normalization. This effect is only present for multiples of
$2\Lam$ and is completely absent in linear optics. 
Our calculation shows that the spin polarized Fe $d$ bands are
responsible for the occurrence and amplification of the observed
oscillations. This becomes apparent if we compare the
resulting SHG intensity of the x-Cu/1Fe/Cu(001) system for opposite
magnetization directions with the system without Fe 
interlayer (but keeping the confinement for the Cu overlayer), which
is 50 times weaker, in good agreement with 
experiment \cite{straub95}. This strong enhancement is
caused by the 
additional terms to be summed for the calculation of \cw when 
more bands are present, even if they are not resonant with the
QWS. This amplification mechanism is not possible in linear
optics, in agreement with experimental observations \cite{suzuki92}.

In Fig.~\ref{M,-M} we show results of our calculation of the NOLIMOKE
signal demonstrating the pronounced QWS oscillations and their strong
spin dependence. The large peak at approx. 11 ML (and a corresponding
peak at 22 ML) results from the amplification due to the  
Fe bands, while the resonant $2\hbar\omega$ transition is between Cu
$d$ and Cu $s$ 
bands (transition (a) in Fig.~\ref{band}). Since the position of the
Fe bands is less important for such a 
constellation, this peak is dominating the SHG spectrum for both
magnetization directions. At 6-7 ML the Cu $d$ band edge is too
far below $E_F$ to give resonances with the QWS for the photon energy
of 1.61 eV, hence a much reduced intensity results. 
While for the majority spins both $\hbar\omega$ resonances 
with Fe $d$ bands as intermediate states and $2\hbar\omega$ resonances 
with Fe as initial state are important, the minority transitions
involve mainly $\hbar\omega$ resonances with intermediate Fe $d$
states and Cu $d$ 
states as initial states. Since these resonances are not well matched
by the photon energy, the short period of the SHG yield from the minority
electrons is less pronounced. This can be traced back to the
$I_{2\omega}(\bf{-M})$ 
yield, which is (in the geometry under consideration) influenced mainly
by the minority transitions. 
The observed slight
 difference of the corresponding periods between the two magnetization
 directions is caused by the exchange splitting of the Fe $d$ bands,
 allowing for resonances with the QWS at different layer numbers. 
The inset of
Fig.~\ref{M,-M} showing results for the magnetic contrast 
$\Delta I_{2w}$ 
gives further evidence for the importance of the Fe $d$ bands. The result
indicates clearly that 
the exchange splitting of the Fe interlayer is involved. 
The contrast varies between $100 \%$ and
$-80\%$ and changes sign several times, due to the same magnitude of the SHG intensity for both
magnetization directions. This
coincidence of the two intensities at fixed frequency could not be
explained if the 
SHG yield would be generated solely by transitions between three
spin polarized quantum well states, since then one signal should be
much more pronounced than the other one. Furthermore, then the signal of a
pure Cu surface should be of the same magnitude as that of the
sandwich. 
Calculations for different exchange
splittings showed that the occurring oscillation periods and the phase
shift between the SHG yield for opposite magnetization direction are strongly
influenced by the strength of the spin splitting. These results
indicate a suppression of periods at particular exchange energies in
sandwich structures.

In Fig.~\ref{moke} we show the dielectric function $\epsilon(\omega)$,
its {\bf M}-dependence and the linear Kerr angle. These results
demonstrate the enhanced sensitivity of NOLIMOKE regarding
oscillations due to QWS as 
compared with the linear optical response. Two oscillation
periods for the imaginary 
part of the dielectric function $\epsilon(\omega)$ for both
magnetization directions can be seen, a dominant one with period 7
ML and a less pronounced oscillation with a period of 3-4 ML. From
Fig.~\ref{band} the origin of these oscillation periods becomes
clear, since for about 6-7 ML (and multiples of these thicknesses)
there are QWS above $E_F$ resonant with a Fe minority $d$ Band. Similarly, for
the other peaks there are respective resonances with Fe $d$
bands. In contrast to nonlinear optics, an overall increase of the 
linear signal with Cu thickness is observed, since it results not
only from the interface, but from all layers, so that the
normalization with respect to the layer thickness has not to be performed.
Of course, the doubled period (11 ML) is absent, since \hhw resonances
do not contribute to the linear signal. Note, the magnetic effect is
three orders of magnitude smaller than for the nonlinear signal, due
to the strong influence of the nonmagnetic intraband
transitions on the linear signal. Thus for the linear susceptibility,
$\chi^{(1)}_{even}(\omega, {\bf M}) \gg \chi^{(1)}_{odd}(\omega,{\bf M})$.
This small magnetic effect becomes obvious from the linear Kerr angle
$\Phi_{\rm Kerr}$ shown in the inset of Fig.~\ref{moke}. $\Phi_{\rm Kerr}$ is
of the order of mdeg, whereas the
nonlinear Kerr angle is two to three orders of magnitude larger
\cite{pusto}. The overall increase with increasing layer thickness 
is again due to the long range of MOKE. The period of 6-7 ML is due to
\hw resonances between QWS above $E_F$ and the Fe majority $d$ band at
-1.4 eV. If this transition is resonant, the majority contribution of
$\epsilon(\omega)$ increases, resulting in an increase of
$\Phi_{\rm Kerr}$ at the corresponding layer thickness. 

In conclusion, we showed that QWS give rise to strongly enhanced
oscillations. The electronic origin of this strong enhancement is
analyzed. We get that NOLIMOKE is able to probe
particular transitions in {\bf k}-space. Our results demonstrate that
although caused by the 
$s$ QWS the amplitude of the oscillation is due to the high density of
Fe $d$ states. Periods different from the fundamental period found in
PE experiments are possible, depending on the position of resonant $d$
bands below $E_F$. In contrast to linear optics, in NOLIMOKE even
$2\hbar\omega$ resonances strongly influence the oscillation. In the
considered sandwich structure, this makes the doubled period to dominate
the spectrum.

Helpful discussions with R. Vollmer, M. Straub, J. Kirschner,
A. Kirilyuk and Th. Rasing are acknowledged. This work was supported
by Deutsche Forschungsgemeinschaft, Sonderforschungsbereich 290.

%
%

\begin{figure}
\noindent
\caption[]{Band structure of the xCu/1Fe/Cu(001) sandwich along
  ${\bf k}^{\perp}$. Bands with energy less than -5 eV are not
  drawn. The numbers 5, 6, 8, 10, 12 on the Fermi energy level indicate the 
  {\bf k}-position of unoccupied QWS as they occur at the corresponding layer
  number. The marked transition (a) has a resonance at
  $2\hbar\omega$ and is responsible for the main peak at approx. 11
  ML. The nonlinear transition (b) is responsible for the
  oscillation period of 6-7 ML. Also the dominating transition for
  MOKE making the 6-7 ML oscillation is indicated (dashed arrow).}
\label{band}
\end{figure}
\begin{figure}
\noindent
\caption[]{SHG yield for opposite magnetization directions {\bf M} and
  {\bf -M}. The dominating 11 ML period is due to a $2\hbar\omega$
  resonance between Cu $d$ states and quantum well states, drastically
  enhanced by the Fe $d$ bands and thus demonstrating the
  {\bf k}-selectivity of NOLIMOKE. The signal for neglected Fe bands is
  nearly vanishing on this intensity scale. The peak shift between the
  {\bf M}- and ({\bf -M})-signal is due to
  the spin polarization of the Fe $d$ bands. $I_{2\omega}$ refers to
  the case where Fe is absent, but the confinement of the Cu layers is
  kept. The inset shows the
  magnetic contrast $\Delta I_{2\omega} = \frac{I_{2\omega}({\bf
      M})-I_{2\omega}({\bf - 
    M})}{I_{2\omega}({\bf M})+I_{2\omega}({\bf - M})}$ . }
\label{M,-M}
\end{figure}
\begin{figure}
\noindent
\caption[]{Linear dielectric function of the x-Cu/1Fe/Cu(001)
  sandwich for opposite magnetization as a function of the Cu layer
  thickness. Note, the 6-7 ML period is visible, while the 11 ML period is completely absent. The magnetic contrast is much  
  smaller than for the NOLIMOKE signal. The inset shows the linear
  Kerr angle as a function of the Cu layer thickness.}
\label{moke}
\end{figure}
%


\begin{references}
\bibitem[32] {footnote}It remains to be shown that contributions which
  in principle could give rise to additional oscillations play a
  significant role.
\end{references}
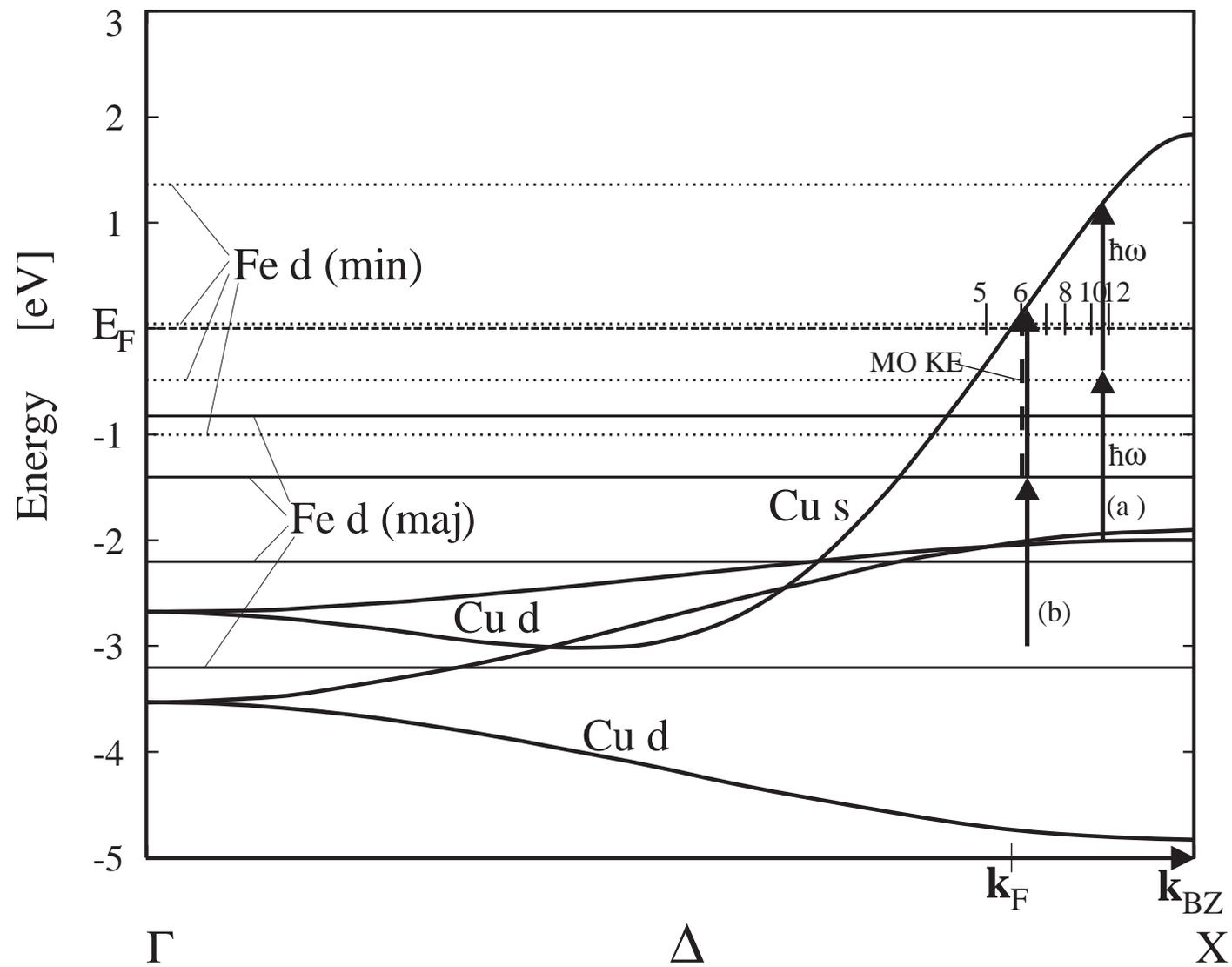
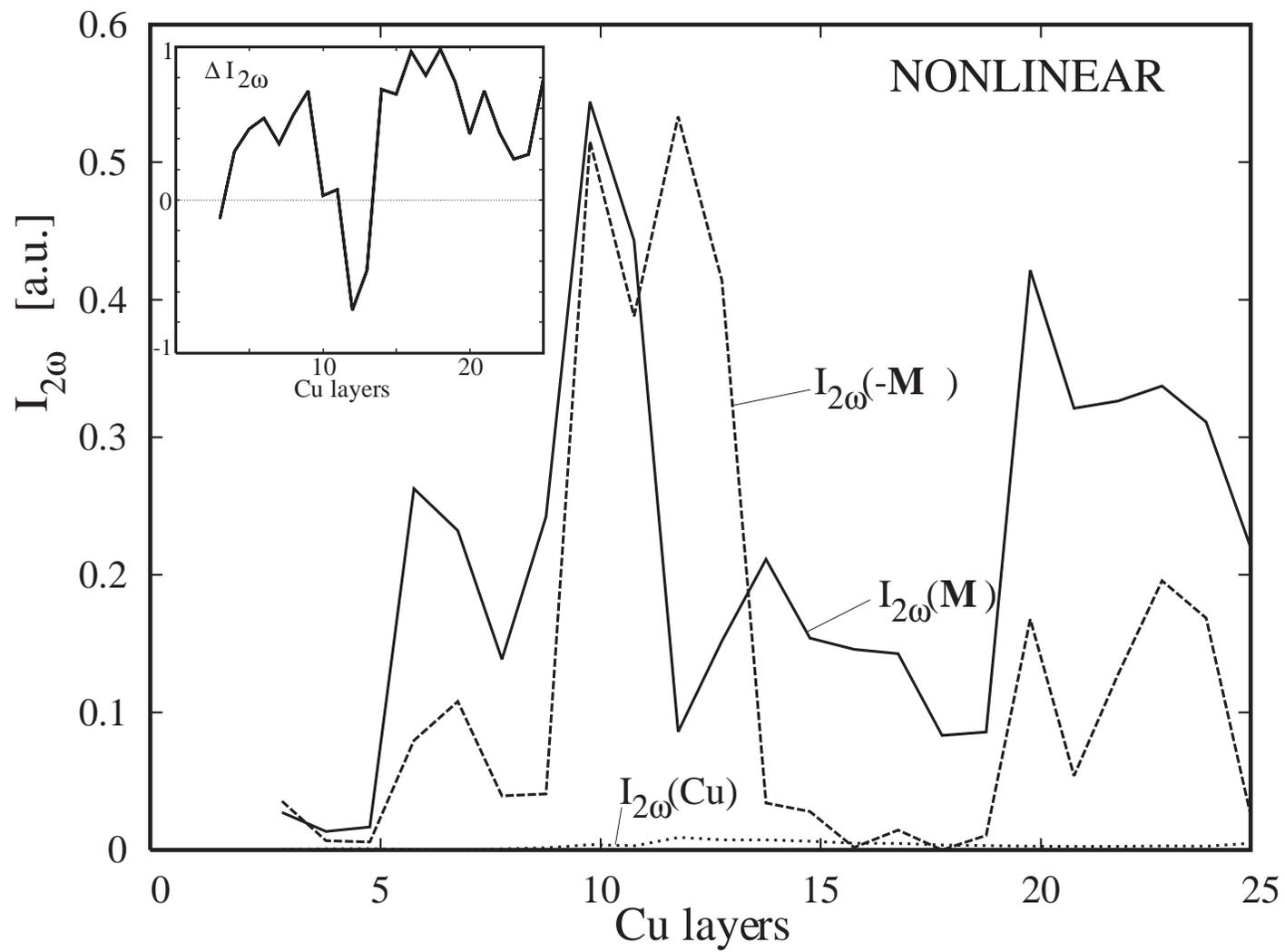
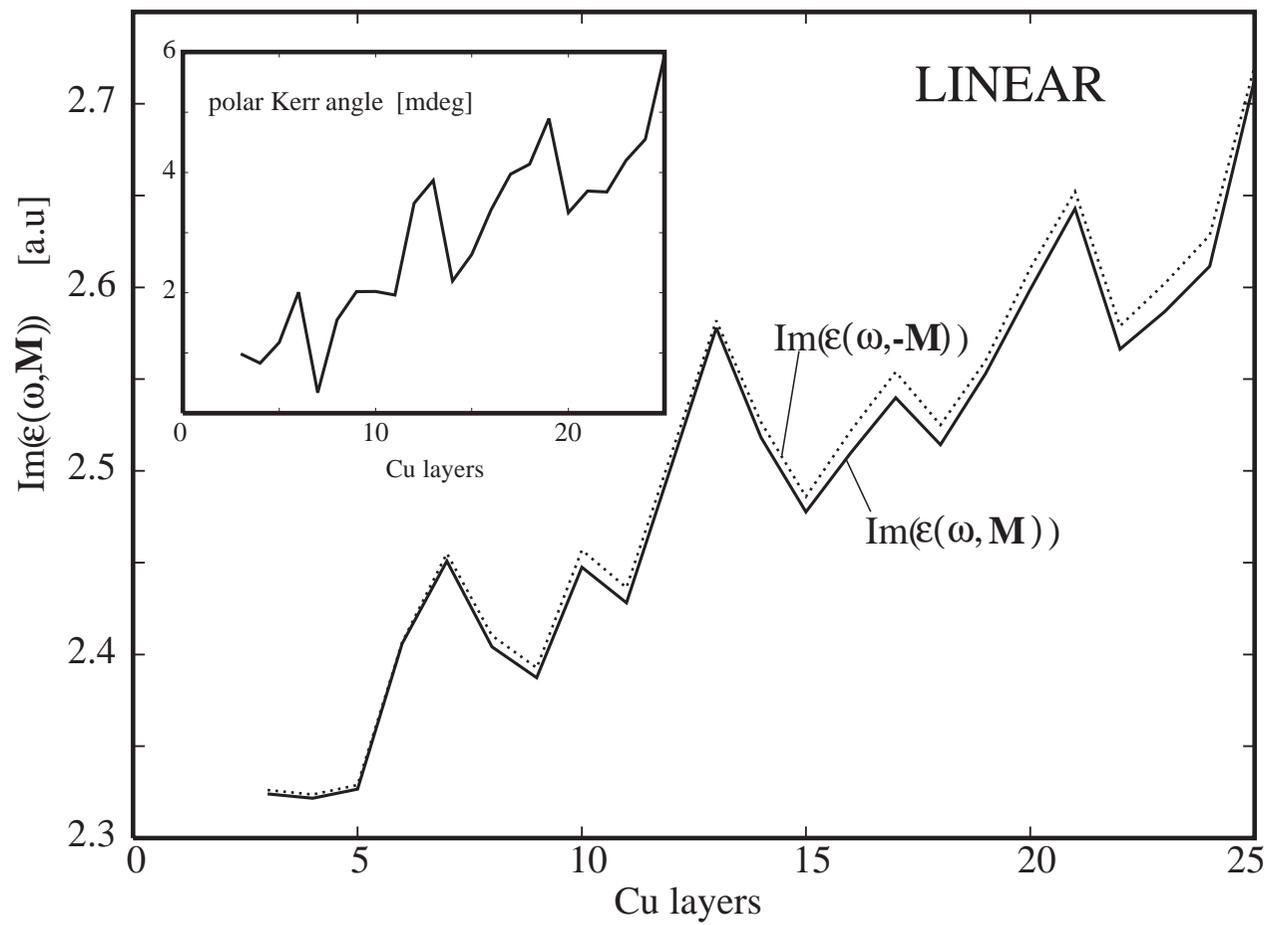
\end{document}